\documentclass[twocolumn,aps,prl,groupedaddress]{revtex4}
\usepackage{graphicx}
\usepackage{color}
\usepackage{amsmath}
\usepackage{marvosym}

\newcommand{\be}{\begin{equation}}
\newcommand{\ee}{\end{equation}}
\newcommand{\bea}{\begin{eqnarray}}
\newcommand{\eea}{\end{eqnarray}}

\newcommand{\la}{\langle}
\newcommand{\ra}{\rangle}

\renewcommand{\vec}[1]{{\bf #1}}
\renewcommand{\epsilon}{\varepsilon}

\renewcommand{\matrix}[1]{\underline{\underline{\boldsymbol{#1}}}}

\begin{document}
\title{Giant Hall photoconductivity in narrow-gapped Dirac materials}
\author{Justin C. W. Song$^{1,2}$} 
\email{justinsong@ntu.edu.sg}
\author{Mikhail A. Kats$^{3,4,5}$}
\affiliation{$^1$Institute of High Performance Computing, Agency for Science, Technology, and Research, Singapore 138632}
\affiliation{$^2$Division of Physics and Applied Physics, Nanyang Technological University, Singapore 637371}
\affiliation{$^3$Department of Electrical and Computer Engineering, University of Wisconsin-Madison, Madison WI 53706, USA}
\affiliation{$^4$Department of Materials Science and Engineering, University of Wisconsin-Madison, Madison WI 53706, USA}
\affiliation{$^5$Department of Physics, University of Wisconsin-Madison, Madison WI 53706, USA}

\begin{abstract}
Carrier dynamics acquire a new character in the presence of Bloch-band Berry curvature, which naturally arises in gapped Dirac materials (GDMs). 
Here we argue that photoresponse in GDMs with small band gaps is dramatically enhanced by Berry curvature. This manifests in a giant and saturable Hall photoconductivity when illuminated by circularly polarized light. 
Unlike Hall motion arising from a Lorentz force in a magnetic field, which impedes longitudinal carrier motion, Hall photoconductivity arising from Berry curvature can boost longitudinal carrier transport. In GDMs, this results in a helicity-dependent photoresponse in the Hall regime, where photoconductivity is dominated by its Hall component. We find that the induced Hall conductivity per incident irradiance is enhanced 
by up to six orders of magnitude when moving from the visible regime (with corresponding band gaps) to the far infrared. These results suggest that narrow-gap GDMs are an ideal test-bed for the unique physics that arise in the presence of Berry curvature, and open a new avenue for infrared and terahertz optoelectronics.
\end{abstract}  

\maketitle

Hall currents
flow in a direction transverse
to an applied electric field. 
This transverse motion, a signature of time-reversal symmetry breaking  
in electronic systems, has its most dramatic impact in the
Hall regime, where the conductivity is dominated by the Hall component. As a result, a number of novel 
behaviors can manifest in DC transport including deformed current flows~\cite{davies}, and a quantized Hall effect~\cite{vonklitzing}. Of current special interest is the dynamical behavior of electronic systems which possess topological Bloch-bands~\cite{essin,qi,armitage2016,tokura2016,molenkamp2016}. In these systems, electrons can feature unusual dynamics that give rise to novel optical and opto-electronic responses, e.g., quantum Faraday and Kerr rotation in the Hall regime~\cite{armitage2016,tokura2016,molenkamp2016}. So far, attaining novel optical/opto-electronic responses associated with topological Bloch-bands have required either large magnetic fields~\cite{armitage2016,molenkamp2016}, or magnetic topological insulators~\cite{tokura2016}.

\begin{figure}
\includegraphics[width=\columnwidth]
{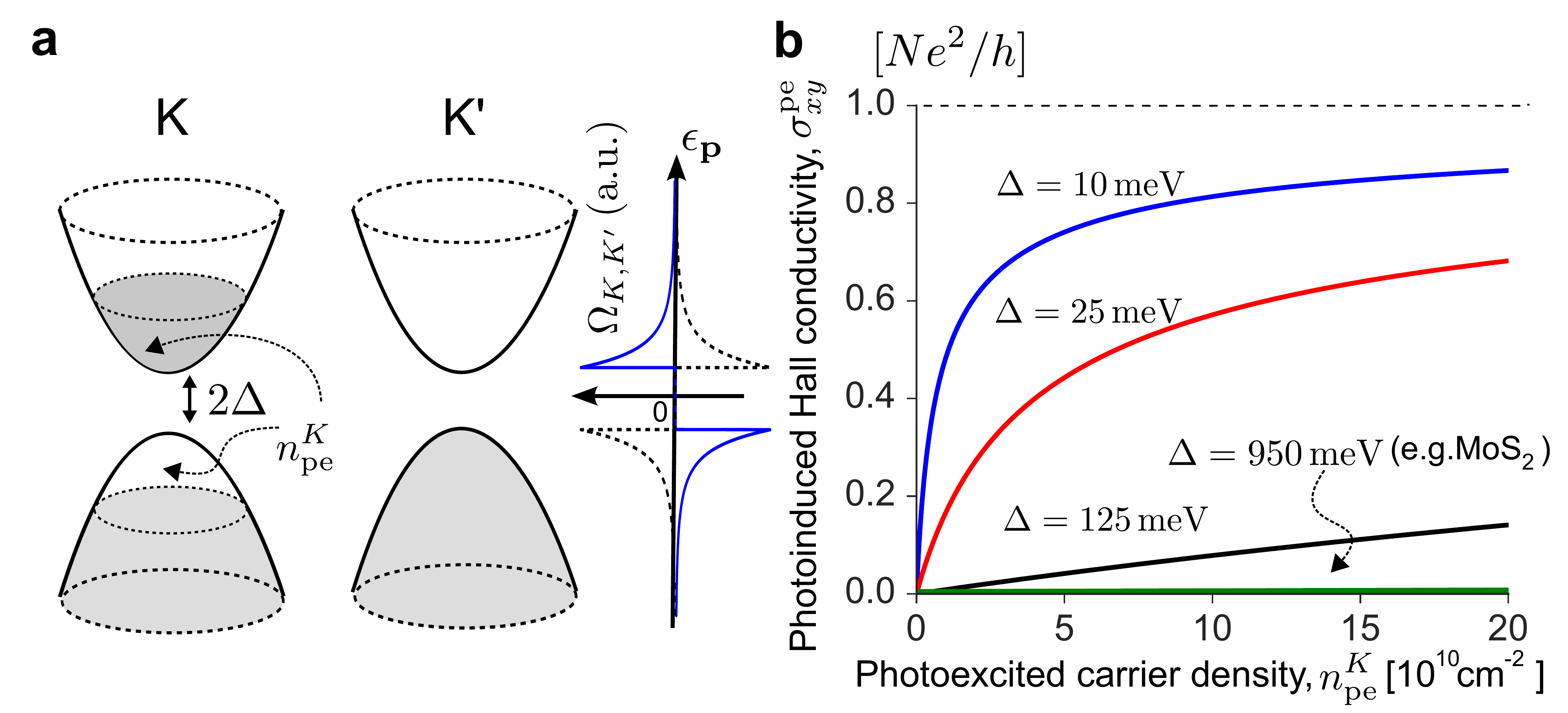}
\caption{ {\bf a)} Population imbalance between valley $K$ and $K'$ in a gapped Dirac material (GDM) with energy gap $2\Delta$. The imbalance can be achieved via absorption of circularly polarized light. [inset] Berry curvature $\Omega(\vec p)$ distribution in valley $K$ (blue) and $K'$ (dashed). {\bf b)} Giant photoinduced Hall conductivity, $\sigma_{xy}^{\rm pe}$, 
as a function of photoexcited carrier density in a single valley, $n^{K}_{\rm pe}$. Note that the green curve, $\Delta = 950 \, {\rm meV}$ (corresponding to MoS$_2$), is almost flat in comparison with those for  narrow-gap GDMs. 
Parameters used: $n_{\rm pe}^{K'} =0$, initial carrier concentration $n_0 = 0$, and $v = 10^{8} \, {\rm cm}{\rm s}^{-1}$. } 
\label{fig:sigmaxy}
\end{figure}

Recently, gapped Dirac materials (GDMs), characterized by an energy gap $2\Delta$~\cite{mak2014,gorbachev2014,tarucha2015,zhang2015,long2015,hunt2013,zhang2009,woods,chen2014,mak2012}, have been found to exhibit a valley Hall effect~\cite{gorbachev2014,tarucha2015,zhang2015,dixiao2007,yao2008,yao2012,mak2014,niureview,lensky2015}. GDMs include transition metal dichalcogenides~\cite{mak2012,mak2014}, dual-gated bilayer graphene (BLG)~\cite{zhang2009,zhang2015,tarucha2015}, and graphene on hexagonal boron nitride heterostructures (G/hBN) with broken A/B sublattice symmetry~\cite{hunt2013,gorbachev2014,woods,chen2014}. In these, valley-specific Hall motion arises at zero magnetic field due to Bloch-band Berry curvature~\cite{dixiao2007,yao2008,yao2012,niureview,lensky2015}. When the valley carrier population is imbalanced (``valley polarized'', see Fig. 1a), a charge Hall effect manifests~\cite{dixiao2007,yao2008,yao2012,mak2014,niureview,lensky2015}. For example, a small Hall voltage at zero magnetic field was observed in monolayer molybdenum disulfide (MoS$_2$), a wide-gap GDM with $2\Delta = 1.9 \, {\rm eV}$~\cite{mak2012}, when valley polarized~\cite{mak2014}. However, the corresponding Hall conductivities measured in Ref.~\onlinecite{mak2014} were small, and the system was far from the Hall regime.

Here we argue that {\it narrow-gap} GDMs
possess a giant photoinduced Hall conductivity $\sigma_{xy}^{\rm pe}$ when the valley carrier populations are imbalanced (Fig.~\ref{fig:sigmaxy}a), which can be achieved via the absorption of light with non-zero helicity~\cite{yao2008}. In particular, we find $\sigma_{xy}^{\rm pe}$ in these narrow-gap GDMs can reach values as high as $N e^2/h$, (Fig.~\ref{fig:sigmaxy}b) even at small photoexcited carrier densities in a single valley $n_{\rm pe}^K$; here $N$ is the spin degeneracy. This contrasts strongly with the case of wide-gap GDMs (Fig. \ref{fig:sigmaxy}b, blue vs. green curves).  
Narrow-gap GDMs can be realized in G/hBN heterostructures and dual-gated BLG, where the gap can take on small values $2\Delta \lesssim {\rm tens} \times {\rm meV}$~\cite{hunt2013,gorbachev2014,tarucha2015,zhang2015,long2015,zhang2009}. 

Key to attaining giant values of $\sigma_{xy}^{\rm pe}$ is 
the $\Delta$ dependence of the anomalous velocity of carriers, $\vec v_a^\zeta$, that characterize their valley Hall motion~\cite{gorbachev2014,tarucha2015,zhang2015,dixiao2007,yao2008,yao2012,mak2014,niureview,lensky2015}. In graphene with broken A/B sublattice symmetry, $\vec v_a^\zeta$ peaks near band edges as~\cite{lensky2015,niureview}
\be
\vec v_a^\zeta = \frac{q}{\hbar} \vec E 
 \times \Omega^{\zeta}_{\pm} (\vec p) \hat{{\vec z}}, \quad 
\Omega_{\pm}^\zeta (\vec p) = 
 \frac{ \pm\hbar^2 \zeta v^2 \Delta}{2(v^2 |\vec p|^2 + \Delta^2)^{3/2}}, 
\label{eq:berry}
\ee
where $\vec E$ is the applied electric field, $q$ is the carrier charge, $\Omega_{\pm}^\zeta (\vec p) $ is the Berry curvature, $\zeta = \pm 1 $ denotes valley $K$ or $K'$, 
$ \pm $ denotes the conduction and valence bands, $v$ is the Dirac particle velocity, and $\vec p$ 
are momenta taken relative to the $K$ and the $K'$ points. 
Importantly, as we explain below, $\Omega (\vec p)$ (see Fig. 1a) exhibits a peak height (width) that gets higher (narrower) as $\Delta$ decreases. As a result, a small $n_{\rm pe}^K$ in narrow-gapped GDMs can carry a large anomalous Hall current.

\begin{figure}
\hspace{-0.2mm}
\includegraphics[width=\columnwidth]{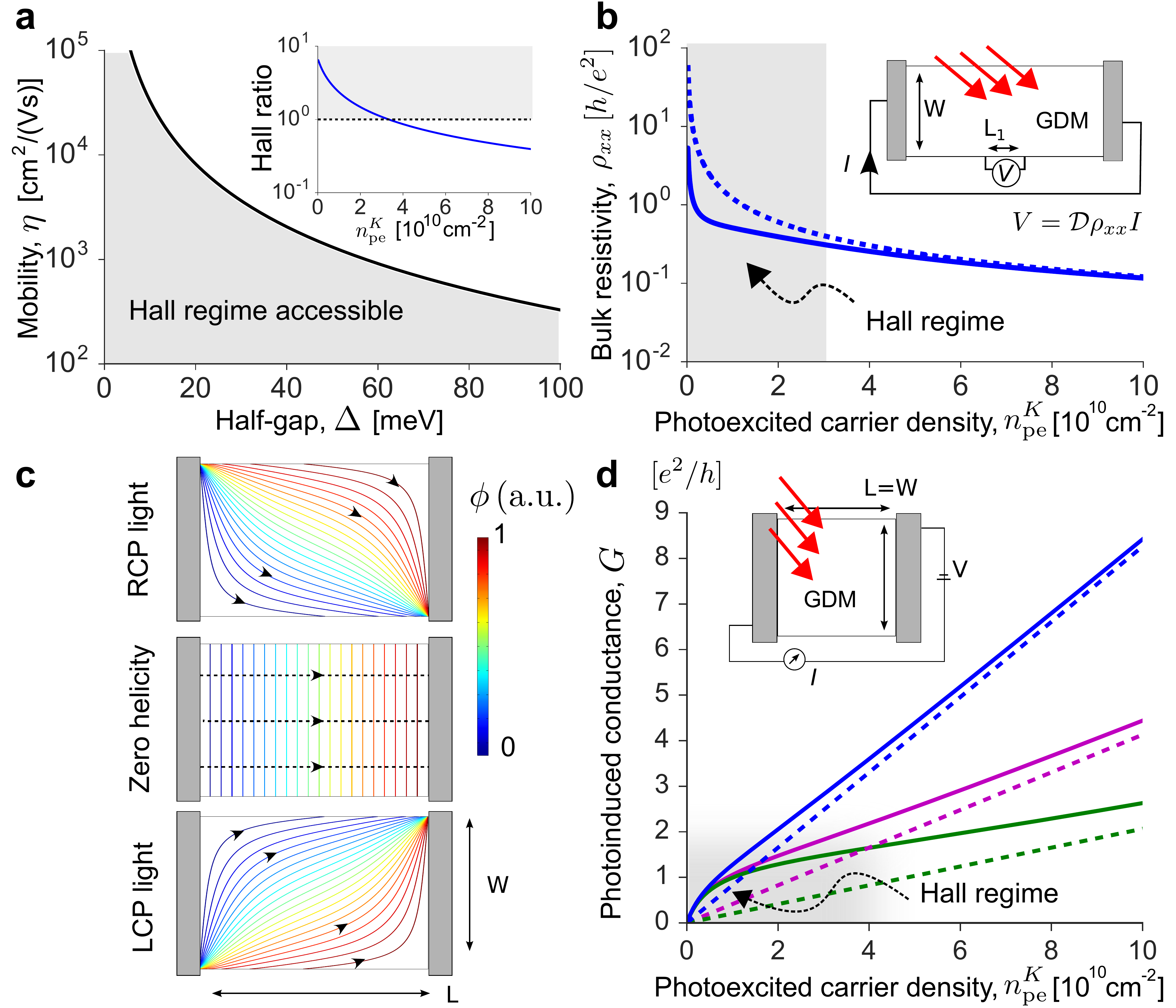}
\caption{ {\bf a)} The Hall regime in GDMs can be accessed when $\sigma_{xy}^{\rm pe} > \sigma_{xx}^{\rm pe}$, indicated by the shaded area which shows the values of $\Delta$ and $\eta$ required in order to achieve the Hall regime. 
The black line corresponds to $\mathcal{A} = 1$ [Eq.~(\ref{eq:hallangle})], where $\mathcal{A}$ is the maximal Hall ratio. 
(inset) The Hall ratio can be controlled by $n_{\rm pe}^K$. Here we plot the Hall ratio for $\Delta = 10 \, {\rm meV}$ and $\eta = 10^4 \, {\rm cm}^2 / ({\rm Vs})$, with the shaded region indicating the Hall regime. {\bf b)} In the Hall regime, longitudinal (bulk) photoresistivity, $\rho_{xx} = V/(I\mathcal{D})$, experiences a large suppression. Solid line parameters used are the same as in the inset to (a). Here $\mathcal{D} = L_1/W$ is a geometrical factor, and the gray region indicates Hall regime. For comparison, the dashed line shows the photoresistivity for $\sigma_{xy}^{\rm pe} =0$.
{\bf c)} Equipotentials in a finite rectangular geometry (length and width, $L$ and $W$) are deformed in the presence of large $\sigma_{xy}^{\rm pe}$, shown for $\sigma_{xy}/\sigma_{xx} = -10, 0, 10$ respectively. Current flow lines (black arrows) are normal to equipotentials when $\sigma_{xy}/\sigma_{xx} = 0$ (center panel). In contrast, in the Hall regime, current flow lines follow the equipotentials resulting from the absorption of LCP/RCP light in GDMs (top/bottom panels, see text).   
{\bf d)} Enhanced two terminal photoconductance, $G = I/V$, for a square geometry $L=W$ (inset), which exhibits a non-linear $n_{\rm pe}^K$ dependence in the Hall regime (gray region). Here $\eta = 1, \, 0.5, \, 0.25  \times 10^{4}  \, {\rm cm}^2/{\rm Vs}$ for blue, magenta, and green curves respectively, with parameters same as (a) inset. Dashed lines correspond to $G$ with $\sigma_{xy}^{\rm pe} = 0$.
The above analysis is valid in the semiclassical limit, where the mean free path is larger than the electron wavelength. }
\label{fig:conductance}
\end{figure}

Strikingly, $\sigma_{xy}^{\rm pe}$ in narrow-gap GDMs can exceed the longitudinal photoinduced conductivity, $\sigma_{xx}^{\rm pe}$, providing access to the Hall regime.  For GDMs, we 
find Hall ratios, ${\rm tan} \, \theta_{\rm H}= \sigma_{xy}^{\rm pe}/\sigma_{xx}^{\rm pe}$, given by
\be
{\rm tan}\, \theta_{\rm H} = \frac{4\mathcal{A} \tilde{n}}{n_{\rm pe}} \Bigg[   \frac{\tilde{n}^{1/2}}{\sqrt{\tilde{n}+ \frac{n_{\rm pe}^{K'}}{2}}}- \frac{\tilde{n}^{1/2}}{\sqrt{\tilde{n}+ \frac{n_{\rm pe}^K}{2}}}\Bigg], 
\quad \mathcal{A} = \frac{e}{4h \tilde{n} \eta}, 
\label{eq:hallangle}
\ee
where $\mathcal{A}$ is the maximal Hall ratio, $n_{\rm pe} = n_{\rm pe}^K + n_{\rm pe}^{K'}$ is the total photoexcited carrier density (per spin), $\tilde{n}= \Delta^2/{4\pi v^2 \hbar^2}$ 
is a gap-dependent characteristic density scale (for $\Delta \neq 0$),
$\eta$ is the mobility of electrons and holes, and $h$ is Planck's constant. 
When the GDM is maximally polarized, e.g. $n_{\rm pe}^K \neq n_{\rm pe}^{K'} = 0$, ${\rm tan} \theta_{\rm H}$ approaches $\mathcal{A}$ for small values of $n_{\rm pe}^K$. As a result, $\sigma_{xy}^{\rm pe} > \sigma_{xx}^{\rm pe}$ in narrow-gap GDMs for a fairly large range of mobilities $\eta$ (shaded region, Fig.~\ref{fig:conductance}a). 

We emphasize that anomalous transport described by 
Eq.~(\ref{eq:berry}) contrasts starkly with that achieved in a magnetic field, with important qualitative consequences. 
In a magnetic field, Hall motion arises from a Lorentz force, which impedes the longitudinal motion of charge carriers. In contrast, $\sigma_{xy}^{\rm pe}$ can enhance longitudinal transport. As we show below, $\sigma_{xy}^{\rm pe}$ suppresses the bulk longitudinal photoresistivity (Fig.~\ref{fig:conductance}b), and enhances the global two-terminal photoconductance (Fig.~\ref{fig:conductance}d). This underscores the dichotomy between magneto-transport, and anomalous transport arising from Bloch-band Berry curvature.

\vspace{2mm}
\section{\bf Hall photoconductivity in GDMs}
We begin with the simplified GDM hamiltonian $\mathcal{H} = H_K + H_{K'}$, describing the low energy excitations in graphene with broken A/B sublattice symmetry~\cite{bernevig} (e.g. G/hBN):
\be
H_{K,K'} = \boldsymbol{\tau} \cdot \vec{d}(\vec p), \quad \vec{d} (\vec p) = ( vp_x, \zeta vp_y , \Delta  )
\label{eq:hamiltonian}
\ee 
where $H_K, H_{K'}$ describe electronic states in the valleys close to the $K$ and $K'$ points, and $\boldsymbol{\tau} = \tau_x \hat{\vec x} +\tau_y \hat{\vec y} + \tau_z\hat{\vec z}$. Here $\tau_{x,y,z}$ are the Pauli matrices operating on the sub-lattices, and $\vec p = (p_x,p_y)$ lies in $x$-$y$ (in-plane). $2\Delta$ is the difference between the on-site energies for the A and B sub-lattices (yielding the gap $2\Delta$) and $v$ is the velocity of Dirac particles for $\epsilon_\vec p \gg \Delta$. We note that Eq.~(\ref{eq:hamiltonian}) captures the essential long-wavelength physics of a broad range of GDMs. In particular, a similar analysis also applies for dual-gated BLG (see Supporting Information). 

The eigenfunctions of Eq.~(\ref{eq:hamiltonian}) are the pseudo-spinors: $| \psi_+^{(0)} (\vec p) \ra = ({\rm cos} \frac{\theta}{2} e^{-\zeta i \phi}, {\rm sin} \frac{\theta}{2})$, and  $| \psi_-^{(0)} (\vec p) \ra = ({\rm sin} \frac{\theta}{2}e^{- \zeta i \phi},-{\rm cos} \frac{\theta}{2})$~\cite{bernevig}, where the two components refer to the wavefunction weight on A and B sites respectively, ${\rm tan} \theta  = v|\vec p| /\Delta$, and ${\rm tan} \phi = p_y/p_x$. These yield energy eigenvalues $\epsilon_{\vec p}^\pm = \pm \sqrt{v^2|\vec p|^2 + \Delta^2}$.

In GDMs, the velocity of an electron wavepacket, $\dot{\vec x}_{\vec p}^\zeta$, in valley $K$ or $K'$ depends on the relative amplitude and phase of the electronic wavefunction on the sub-lattice A and B sites. For example, when the wavefunction weight is completely on the A site, the carrier velocity vanishes; velocity similarly vanishes when the wavefunction weight is on the B site only. Using the wavefunction weight on A and B sub-lattice sites shown in $|\psi_\pm^{(0)}(\vec p)\ra$ above, the carriers in the $\pm$ bands possess the usual group velocity $\partial{\epsilon}_\vec p^\pm/{\partial \vec p}$.

However when an electric field $\vec E$ is applied, the wavefunction becomes perturbed $|\psi_\pm (\vec p) \ra = |\psi_\pm^{(0)}(\vec p) \ra + |\delta \psi (\vec p) \ra$, altering the wavefunction amplitude and phase on the A and B sub-lattice sites.  
As a result, $\dot{\vec x}_{\vec p}^\zeta$ acquires an additional contribution, the anomalous velocity $\vec v_a^\zeta$, as~\cite{niureview} 
\be
 \dot{\vec x}_{\vec p}^{\zeta} = \frac{\partial \epsilon^\pm_{\vec p}}{\partial \vec p} + \vec v_a^\zeta, 
\label{eq:vel}
\ee
where anomalous velocity is given in Eq.~(\ref{eq:berry}); the energy dependence of $\vec v_a^\zeta$ in Eq.~\ref{eq:berry} can be obtained 
from a perturbative analysis, see e.g. Section 2 in Ref.~\onlinecite{niureview}. Intuitively, we note that first order perturbation theory in $\vec E$ 
yields $|\delta \psi (\vec p) \ra$ that scales inversely with the energy difference between the $+$ and $-$ states, ${\epsilon}_\vec p^+ - {\epsilon}_\vec p^- $. Therefore $|\delta \psi (\vec p) \ra$ and $\vec v_a^\zeta$ are peaked at the band edge where the energy difference reaches a minimum of $2\Delta$ (see Fig.~\ref{fig:sigmaxy}a inset). Importantly, for carriers close to the band edge ($v|\vec p|\ll \Delta$), the anomalous velocity $\vec v_a^\zeta \propto 1/\Delta^2$[Eq.~(\ref{eq:berry})]. As a result, for $2\Delta$ small in narrow-gap GDMs, band edge carriers attain giant values of $\vec v_a^\zeta$, enabling them to carry a large anomalous Hall current.

The anomalous velocity, $\vec v_a^\zeta$, skews the electron velocity $ \dot{\vec x}_{\vec p}^\zeta$ so that it is no longer parallel to its momentum. As a result, Bloch electrons in each of the valleys experience a Hall effect even without the application of a magnetic field~\cite{niureview,nagaosareview}, albeit of opposite sign. In our analysis below, we do not consider extrinsic disorder-related mechanisms which can contribute to the anomalous Hall effect~\cite{nagaosareview}. Instead, we concentrate on the intrinsic mechanism in GDMs, arising from $\vec v_a^\zeta$, to clearly illustrate the features in the photo-induced Hall regime.

We proceed by writing the current density $\vec j = q\sum_{\vec{p},\pm,\zeta} f_{\pm}^\zeta (\vec p) \dot{\vec x}$, yielding $ j_i = \sigma_{ij} E_j$. 
Here $f_\pm^\zeta (\vec p)$ 
is the 
distribution function of the carriers with 
$\pm, \zeta$ denoting conduction/valence bands and valley $K/K'$ respectively. Focussing on the Hall conductivity, and using Eq.~(\ref{eq:vel}), we have 
\be
\sigma_{xy} = \frac{Ne^2}{\hbar}\Big[ \sum_{\vec{p},\pm} f_{\pm}^K(\vec p) \Omega_{\pm}^{K}(\vec p) +  \sum_{\vec{p},\pm} f_{\pm}^{K'}(\vec p) \Omega_{\pm}^{K'}(\vec p) \Big]. 
\label{eq:general}
\ee
We note that at zero magnetic field, $\Omega_{\pm}^{K'} = -\Omega_{\pm}^{K}$~\cite{niureview}.  At equilibrium, the distribution functions $f_{K,\pm} = f_{K',\pm}$, giving equal carrier densities in each of the valleys, $n_K = n_{K'}$, and, as a result, $\sigma_{xy}$ vanishes. However, when pushed out of equilibrium, $f_{K} \neq f_{K'}$, yielding $\sigma_{xy} \neq 0$.

Assuming fast relaxation of the photoexcited carriers to the band-edge~\cite{gierz2013,johannsen2013}, the steady-state carrier population in each of the valleys can be modeled by distribution functions with split electron and hole quasi-Fermi levels:
\be
f_+^\zeta (\vec p) = [1+ e^{\beta (\epsilon_{\vec p}^+ - \mu_{{\rm el},\zeta})}]^{-1}, \quad f_-^\zeta (\vec p) = [1+ e^{\beta (\epsilon_{\vec p}^- - \mu_{{\rm h},\zeta})}]^{-1}, 
\label{eq:distributionfunction}
\ee
where $\mu_{\rm el,\zeta} \neq \mu_{\rm h,\zeta}$ are the electron and hole quasi-fermi levels induced by photoexcitation, $\beta = 1/k_B T$, and $T$ the temperature.

\begin{figure}
\includegraphics[width=\columnwidth]{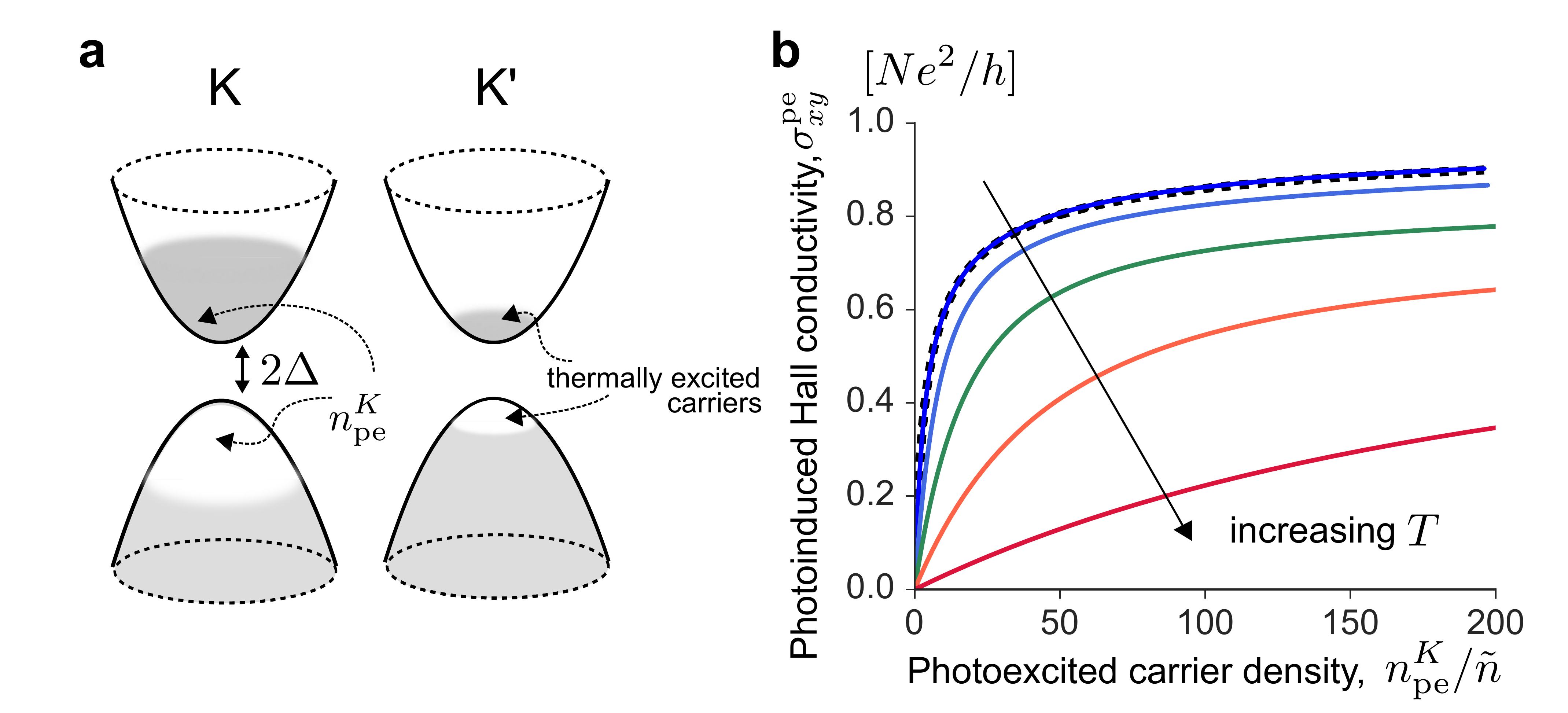}
\caption{ a) At finite temperature, both $K$ and $K'$ possess thermally excited carriers. b) Photoinduced Hall conductivity, $\sigma_{xy}^{\rm pe}$, decreases as a function of temperature. Shown are plots obtained numerically integrating Eq.~(\ref{eq:general}) for $k_BT/\Delta = 0.1,0.5,1,2,5$ (top to bottom); dashed curve is Eq.~(\ref{eq:sigmaxy}) where $T=0$ and we have set the chemical potential before photoexciation to be in the gap. Note that $k_B T/\Delta = 2.6$ for room temperature and $\Delta =10\, {\rm meV}$.}
\label{fig:temp}
\end{figure}

We first examine the degenerate limit, $\mu_{\rm el,h}, \Delta \gg k_BT$. 
Writing the initial carrier density per spin {\it before} irradiation as $n_{K}=n_{K'} = n_0$, and using Eq.~(\ref{eq:general}) and Eq.~(\ref{eq:distributionfunction}), we obtain a photo-induced Hall conductivity as 
\begin{align}
\sigma_{xy}^{\rm pe} &= \frac{N e^2}{h} \Big[\mathcal{F}_{K} (n_{\rm pe}^K/2) + \mathcal{F}_{K'}(n_{\rm pe}^{K'}/2) \Big], \nonumber \\ 
\mathcal{F}_{\zeta}(x) &=  \frac{\zeta}{2} \Big(1- \Big[\frac{\tilde{n}^{1/2}}{\sqrt{\tilde{n} + n_0 + x}} + \frac{\tilde{n}^{1/2}}{\sqrt{\tilde{n} + x}}\Big]  \Big),
\label{eq:sigmaxy}
\end{align}
where $\mathcal{F}_{\zeta}$ with $\zeta = \pm 1$ denote contributions of the $K$ and $K'$ valleys respectively, and $\tilde{n}= \Delta^2/{4\pi v^2 \hbar^2}$ is the characteristic density. Here we noted that the photoexcited electron and hole densities can be written as $n_{\rm el,h} = \mu_{\rm el,h}^2/ 4\pi \hbar^2 v^2 - \tilde{n}$, and neglected finite $T$ corrections by setting $T=0$ (see below for discussion of temperature dependence). We have also taken $n_{\rm el}^{K} = n_{\rm h}^{K} = n_{\rm pe}^{K}/2$, and $n_{\rm el}^{K'} = n_{\rm h}^{K'} = n_{\rm pe}^{K'}/2$, where $n_{\rm el}$ and $n_{\rm h}$ are the photoexcited electron and hole populations in the $K,K'$ valleys. 

Crucially, $\tilde{n}= \Delta^2/{4\pi v^2 \hbar^2}$ determines the characteristic scale with which valley population imbalance, $\delta n = n^{K}_{\rm pe} - n^{K'}_{\rm pe}$, yields appreciable $\sigma_{xy}$ (of the order $N e^2/h$). For example, for $\Delta = 10 \, {\rm meV}$, we find $\tilde n  = 1.8  \times 10^{9} \, {\rm cm}^{-2}$ (setting $v = 10^8 \, {\rm cm}\, s^{-1}$). As a result in narrow-gap GDMs, only a small imbalance $\delta n$ is needed to achieve sizable $\sigma_{xy}^{\rm pe}$. This unusual behavior is a result of states close to the band edge contributing the maximum amount to Hall conductivity in GDMs~\cite{lensky2015} (Fig.~\ref{fig:sigmaxy}a). 

Taking $n_0 =0$, we find a giant photoinduced Hall conductivity $\sigma_{xy}^{\rm pe}$, shown in Fig.~\ref{fig:sigmaxy}b. Here we have set $n_{\rm pe}^{K'} =0$, with only carriers in the $K$ valley excited. The maximum achievable $\sigma_{xy}^{\rm pe}$ is  $N e^2/h$ as shown in Eq.~(\ref{eq:sigmaxy}), when $n_{\rm pe}^{K} \gg \tilde{n}$. Since $\tilde{n}$ can be small in GDMs with narrow gaps, this saturation behavior can be easily accessed. An opposite sign of Hall conductivity is obtained when instead $K'$ carriers are photoexcited, e.g., by absorption of light with opposite helicity. $\sigma_{xy}^{\rm pe}$ can be directly measured in a four-terminal Hall-type geometry, as in Ref.~\onlinecite{mak2014}. 

Interestingly, $\sigma_{xy}^{\rm pe}$ [Eq.~(\ref{eq:sigmaxy})] can be tuned by gating the material, and exhibits an {\it even} behavior in $n_0$ reaching a maximum when $n_0 =0$ (charge neutrality). Due to its even-ness, $\sigma_{xy}^{\rm pe}$ is robust to density inhomogeneity. This
contrasts with that of ordinary Hall conductivity, which changes sign with carrier type, and vanishes for charge-neutral systems. 

At finite temperature, both $K$ and $K'$ valleys possess thermally excited carriers (Fig.~\ref{fig:temp}a). We plot the temperature dependence of $\sigma_{xy}^{\rm pe}$ in Fig.~\ref{fig:temp}b. Here we have integrated Eq.~(\ref{eq:general}) numerically using Eq.~(\ref{eq:distributionfunction}). The photoexcited carrier density was also obtained numerically via $n_{\rm pe}^{K,K'} = \sum_\vec p f_{\zeta,\pm} (\vec p)$. In our plots, we characterized the temperature dependence via the dimensionless parameter $k_BT/\Delta$.
As shown in Fig.~\ref{fig:temp}b, $\sigma_{xy}^{\rm pe}$ {\it decreases} as temperature increases. In particular, temperature dependence of $\sigma_{xy}^{\rm pe}$ becomes pronounced only when $k_BT/ \Delta \gtrsim 1$. This arises since $\Omega(\vec p)$ in Eq.~(\ref{eq:berry}) only varies appreciably when $v |\vec p| \sim \Delta$. As $T$ is increased, the typical square-root type dependence of $\sigma_{xy}^{\rm pe}$ on $n_{\rm pe}^K$ straightens out, approaching a nearly linear dependence for large temperatures (see e.g. $k_BT/\Delta  = 5$ in Fig.~\ref{fig:temp}b). In plotting Fig.~\ref{fig:temp}b, we have chosen the chemical potential before photoexcitation to lie inside the gap, so that $n_0 = 0$ for $T=0$.  

\vspace{2mm}
\section{Photoinduced Hall regime}
To give a sense of scale, it is useful to compare $\sigma_{xy}^{\rm pe}$ to the longitudinal photoconductivity, $\sigma_{xx}^{\rm pe}$. We use a simple model for $\sigma_{xx}^{\rm pe}= e N \eta n_{\rm pe}$, where $n_{\rm pe} = n_{\rm pe}^K + n_{\rm pe}^{K'}$ is the photoexcited carrier density (per flavor/spin $N$), and $n_0 =0$. Here $\eta$ is the mobility of electrons and holes. Taking the ratio of $\sigma_{xy}^{\rm pe}$ in Eq.~(\ref{eq:sigmaxy}) and $\sigma_{xx}^{\rm pe}$, we obtain the Hall ratio in Eq.~(\ref{eq:hallangle}). 
Setting the maximal Hall ratio to unity, $\mathcal{A} = 1$, we obtain the black curve in Fig.~\ref{fig:conductance}a. This indicates that a large range of mobilities yield $\sigma_{xy}^{\rm pe} > \sigma_{xx}^{\rm pe}$, see shaded region. The Hall ratio can be controlled via the photoexcited carrier density, and is maximized at small values of $n_{\rm pe}^K$ (Fig.~\ref{fig:conductance}a inset). We note that when $\Delta =0$, the terms inside the brackets of Eq.~(\ref{eq:hallangle}) vanish.

For G/hBN, gap sizes $2\Delta \approx 10 - 40 \, {\rm meV}$, and mobility varying from $\eta \approx 5 \times 10^3$ to $10^5\, {\rm cm}^{-2}/({\rm Vs})$ have been inferred~\cite{gorbachev2014,hunt2013,chen2014,woods}. We anticipate that for the smallest gap sizes, $\Delta \approx {\rm few} \times {\rm meV}$~\cite{hunt2013,gorbachev2014,chen2014,woods}, and the lowest mobilities, Hall ratios of order ten may be obtained. 

Access to the photoinduced Hall regime invites a comparison with magneto-transport of charge carriers. In the latter,
the Lorentz force due to the presence of a magnetic field impedes the longitudinal motion of charge carriers. For example, for a single carrier type in a homogeneous system, a simple Drude model yields bulk $\sigma_{xx}$ that is suppressed with increasing magnetic field; it also yields a longitudinal resistivity that is unchanged with magnetic field~\cite{davies}. 

In contrast, in the photoinduced Hall regime in GDMs, we find that (bulk) longitudinal {\it photoresistivity}, $\rho_{xx}$, becomes suppressed. Writing $\vec j (\vec r)= \matrix{\sigma} \,\,  [- \boldsymbol{\nabla} \phi (\vec r) ]$, and 
inverting $\matrix{\sigma}$, we obtain 
\be
\rho_{xx} = \frac{\sigma_{xx}}{\sigma_{xx}^2 + \sigma_{xy}^2}= \frac{\mathcal{C}}{\sigma_{xx}^{\rm pe}}, \quad \mathcal{C} = \frac{1}{1 + ({\rm tan} \theta_H)^2},
\label{eq:rhoxx}
\ee
where we have set $\sigma_{xx} =\sigma_{yy} = \sigma_{xx}^{\rm pe}$ to accentuate the contribution of the photoexcited carriers. In the Hall regime in Fig.~\ref{fig:conductance}b, we find $\mathcal{C} \ll 1$ leading to a suppressed $\rho_{xx}$ (solid line) as compared to the case of $\sigma_{xy} = 0$ (dashed line). For example, $\Delta = 10 \, {\rm meV}$ and $\eta = 10^4 \, {\rm cm}^2/(Vs)$, we find $\mathcal{C} \approx 0.1$ resulting in a large suppression. This can be measured via non-invasive local probes in long Hall bars (Fig.~\ref{fig:conductance}b inset), so that the probes are far away from distorted fields near the contacts. 

The dichotomy between magneto-transport and the photoinduced Hall regime in GDMs arises from the differences in the origin of their Hall motion. In magneto-transport, the Lorentz force alters the force balance on charge carriers, and diverts momentum transversely; this impedes longitudinal motion. In the photoinduced Hall regime, $\vec v_a$ appears in addition to longitudinal flow in Eq.~(\ref{eq:vel}), and does not obstruct longitudinal motion since it does not enter into the force balance. As a result, smaller electric fields are sustained giving a suppressed $\rho_{xx}$ in Eq.~(\ref{eq:rhoxx}) \footnote{We note that $\sigma_{xx}^{\rm pe}$ is unaffected by $\vec v_a$ in the photoinduced Hall regime. This contrasts with a single carrier Drude model in a magnetic field, which exhibits a suppressed $\sigma_{xx}$ and constant $\rho_{xx}$. This displays the duality between Berry curvature effects and magneto-transport.}.

The Hall regime can also change the ``global'' pattern of current flow~\cite{davies,lippmann1958,jensen1972,abanin}, as shown in  Fig.~\ref{fig:conductance}c. Here we have contrasted $\phi(\vec r)$ given $|\sigma_{xy}| \gg \sigma_{xx}$ (top and bottom panels) with $\sigma_{xy} = 0$ (middle panel)
\footnote{Here we have solved current continuity $\boldsymbol{\nabla} \cdot \vec j = 0$ in steady state, with boundary conditions $\phi (\vec r)\big|_{\vec r = c_k}= {\rm const}$ and $\vec j (\vec r)\cdot \hat{\vec n} \big|_{\vec r = b_k} = 0$, where $c_k$ and $b_k$ label boundaries with metallic contacts (e.g., source and drain) and without metallic contacts (e.g., sample boundaries) respectively. Here $\hat{\vec n}$ is the unit normal vector to the boundary, and we have used uniform $\sigma$.}. 
In the former case, current flow is non-uniform. Indeed, when the sign of $\sigma_{xy}$ is changed, e.g. by changing between left- (LCP) and right- (RCP) circularly polarized irradiation, the location of ``hot-spot'' regions of largest electric field can switch.

As a result, the two-terminal conductance $G = I/V$, acquires a sensitivity to $\sigma_{xy}$~\cite{lippmann1958,jensen1972,rendell,abanin}. Here $I = \int_0^W j_y dx $, and $V = \phi_{\rm source} - \phi_{\rm drain}$ (Fig.~\ref{fig:conductance}d inset). In an arbitrary geometry, $G$ can be obtained numerically~\cite{rendell,abanin}. However, for a conducting square $L=W$~\cite{lippmann1958,jensen1972},
\be
G = \big[(\sigma_{xy})^2 + (\sigma_{xx})^2\big]^{1/2} = \sigma_{xy}^{\rm pe} \sqrt{1 + 1/({\rm tan}\, \theta_{\rm H})^2},
\label{eq:conductance}
\ee
where in the last line we have set $\sigma_{xx} = \sigma_{xx}^{\rm pe}$, and $\sigma_{xy} = \sigma_{xy}^{\rm pe}$ (realizable for $n_0 =0$ and $k_BT/\Delta \ll 1$). For brevity, we focus on this geometry. Using $\sigma_{xy} = \sigma_{xy}^{\rm pe}$ obtained for GDMs in Eq.~(\ref{eq:sigmaxy}), we obtain the two-terminal photoconductance in Fig.~\ref{fig:conductance}d. 

In the Hall regime at small $n_{\rm pe}^{K}$, $\sigma_{xy}^{\rm pe}$ provides the dominant contribution to the two-terminal conductance $G$ (Fig.~\ref{fig:conductance}d), which   
displays a non-linear $n_{\rm pe}^{K}$ dependence that resembles that of $\sigma_{xy}^{\rm pe}$ 
in Fig.~\ref{fig:sigmaxy}b. This is particularly pronounced at low $n_{\rm pe}^{K}$, where $G$ displays a ``universal'' profile that is insensitive to mobility, boosting $G$. 
At large $n_{\rm pe}^K \gg \tilde{n}$, the Hall ratio diminishes, and $G$ loses its sensitivity to the helicity of the incident light (dashed line). In that regime, $\sigma_{xx}$ dominates $G$.

The boost to $G$ in the photoinduced Hall regime in GDMs contrasts with that of magneto-transport. In the latter, $G$ decreases in the presence of a magnetic field in the semiclassical limit~\footnote{This can be seen by writing for a single carrier Drude model: $\sigma_{xx} = \sigma_0/(1+ \omega_c^2 \tau^2)$, and $\sigma_{xy} = \sigma_0 \omega_c\tau/(1+ \omega_c^2 \tau^2)$, and substituting into Eq.~(\ref{eq:conductance}). Here $\sigma_0$ is longitudinal conductivity at zero magnetic field, $\omega_c$ is the cyclotron frequency, and $\tau$ is the scattering time.}. The dependence of $G$ on the degree of circular polarization, 
and the boost to longitudinal carrier motion [Eq.~(\ref{eq:hallangle}), Eq.~(\ref{eq:conductance})],
are striking signatures of the unique Hall regime accessed in narrow-gap GDMs. 

\vspace{2mm}
\section{\bf Enhanced valley-imbalance rate}
We now turn to valley-imbalance, $\delta n = n^{K}_{\rm pe} - n^{K'}_{\rm pe}$, induced by the absorption of circularly polarized light~\cite{yao2008,mak2014}. As we argue below, narrow-gap GDMs can experience an enhanced rate of imbalance between the valleys $d \delta n/dt$ (as compared with their wide-gap GDM counterparts)
when irradiated by light of non-zero helicity. We model the rate of electron-hole pair creation in each valley, $W_{K(K')}$, 
via Fermi's golden rule
\be
W_{K(K')} = \frac{2\pi }{\hbar} \sum_{\vec k} |M_\vec k^{K (K')} |^2 \delta (\epsilon_\vec k - \hbar \omega/2),
\label{eq:rate-1}
\ee
where the matrix elements are $ M_\vec k^{K (K')}  = \frac{iev}{2\omega} \la \psi_+ | E_x \tau_x + \zeta E_y \tau_y | \psi_- \ra$~\cite{katsnelson}, and the incident light electric field is $\vec E$ with photon energy $\hbar \omega \geq 2 \Delta$. Eq.~(\ref{eq:rate-1}) arises from writing $\vec p \to \vec p - e\vec A/c$ in Eq.~(\ref{eq:hamiltonian}), with the vector potential satisfying $\vec A =  \frac{ic}{\omega} \vec E$. 

Using the pseudo-spinor states $| \psi_{\pm} (\vec p) \ra$ for GDMs given above, we have $\la \psi_+ | \tau_x | \psi_- \ra = {\rm sin}^2 \frac{\theta}{2} e^{-i\zeta \phi} - {\rm cos}^2\frac{\theta}{2} e^{i\zeta  \phi}$, $\la \psi_+ | \tau_y | \psi_- \ra= i\big({\rm sin}^2 \frac{\theta}{2} e^{- i\zeta\phi} + {\rm cos}^2\frac{\theta}{2} e^{ i\zeta \phi}\big)$. For normally incident circularly polarized light, $\vec E = |\vec E| ( \hat{\vec x} + i d\hat{\vec y}) /\sqrt{2}$, where $d=\pm 1$ for LCP and RCP polarizations, which gives the rate (per spin $N$)
\be
W_K^{d} = W_0 \big( \frac{2\Delta}{\hbar \omega} +d\big)^2, \	\quad W_{K'}^{d} = W_0 \big( \frac{2\Delta}{\hbar \omega} -d\big)^2,
\label{eq:rates-2}
\ee
where $W_0 = e^2 |\vec E|^2/(16 \hbar^2 \omega) $, and $\hbar \omega \geq 2 \Delta$. Eq.~(\ref{eq:rates-2}) describes valley selective electron-hole transitions resulting from the absorption of light with non-zero helicity, in agreement with Ref.~\onlinecite{yao2008}. 
 
The $K$ and $K'$ asymmetry in Eq.~(\ref{eq:rates-2}) yields a valley population imbalance rate (per spin $N$)
\be
\frac{d \delta n }{dt } = 2 \big(W_K^{d} - W_{K'}^{d}\big) = d \frac{e^2 |\vec E|^2}{\hbar^2 \omega} \frac{\Delta}{\hbar \omega},
\label{eq:rate-3}
\ee
where $\hbar \omega \geq 2\Delta$, and the factor of $2$ in the second line accounts for both electron and hole populations. The rate $d \delta n/dt $ grows quickly for decreasing $\omega$, and reaches a maximum at $\hbar \omega = 2\Delta$, yielding ${\rm max} ( d \delta n/dt )= e^2 |\vec E|^2/ 4 \hbar \Delta $; $d \delta n/dt $ vanishes for $\hbar \omega < 2\Delta$.  As a result, large $K/K'$ carrier density imbalance rates, $d \delta n/dt$, can be achieved for narrow-gap GDMs when irradiated with circularly polarized light on resonance with the gap. 

Combining the enhancements for narrow-gap GDMs in both Eq.~(\ref{eq:sigmaxy}) and Eq.~(\ref{eq:rate-3}) yields a large value of $\sigma_{xy}^{\rm pe}$ per incident light irradiance, $\mathcal{P}_{\rm in} = c |\mathbf{E}|^2/ (4\pi)$. This is particularly striking in the {\it low} irradiance regime 
where we can expand Eq.~(\ref{eq:sigmaxy}) to lowest order in $\delta n$. This gives a {\it linearized} photoinduced Hall conductivity, $\tilde{\sigma}_{xy}^{\rm pe}$, as
\be
\frac{\tilde{\sigma}_{xy}^{\rm pe}}{ \mathcal{P}_{\rm in}} =  d\frac{S_0}{(\hbar\omega)^2 \Delta}, \quad S_0 = \frac{4N e^2}{h} \frac{ \alpha \pi^2  \hbar^2 v^2}{\gamma}, 
\label{eq:photoconductivity}
\ee
where $\alpha = e^2/\hbar c$ is the fine structure constant, and we have estimated $\delta n = \gamma^{-1}( d \delta n/ dt )$, where $\gamma$ the net relaxation rate that may include electron-hole recombination as well as inter-valley scattering; we have taken $n_0 = 0$. For incident LCP/RCP light with frequency above the gap, $\tilde{\sigma}_{xy}^{\rm pe}/\mathcal{P}_{\rm in}$ scales as $[(\hbar \omega)^2 \Delta]^{-1}$, reaching a maximum for $\hbar \omega = 2\Delta$, where it scales as $\tilde{\sigma}_{xy}^{\rm pe}/\mathcal{P}_{\rm in} \propto 1/\Delta^3$. As a result, we find $\tilde{\sigma}_{xy}^{\rm pe}/\mathcal{P}_{\rm in}$ can vary over six orders of magnitude from $\Delta \approx 1\, {\rm eV}$ (e.g. MoS$_2$) to $\Delta \approx 10 {\rm meV}$ (e.g. G/hBN). This large scaling clearly underscores how the effect of Berry curvature on photoresponse is maximized for narrow-gap GDMs.

In conclusion, GDMs with small gaps are an ideal system in which to probe and utilize the anomalous motion of carriers resulting from the presence of Berry curvature. The giant Hall photoconductivity in such narrow-gap GDMs results in unique phenomena that can be observed at room temperature, including access to the Hall regime in the absence of a magnetic field. Unlike the case of the conventional Hall effect due to a magnetic field, which impedes longitudinal motion, Hall photoconductivity arising from Berry curvature in GDMs boosts it. This demonstrates the contrast between conventional magneto-transport, and Berry curvature induced transport, and unveils new means of controlling carrier transport at zero magnetic field. 

The implications for far-infrared and terahertz optoelectronics are intriguing. While the Hall photoresponse of GDMs with large gaps -- e.g. molybdenum disulfide (MoS$_2$) -- is likely too small for practical optoelectronics applications, the corresponding photoresponse in narrow-gap GDMs, such as graphene on hexagonal boron nitride (G/hBN), can be significant. In a simple two-terminal photoconductor, the anomalous conductance boost can provide additional sensitivity at low input powers. Additionally, combining such a valley Hall photoconducting detector with a traditional photodetector can yield a measurement of the degree of circular polarization of an incident wave, without the need for wave-plates or other polarization-conversion schemes. Furthermore, direct measurements of the Hall photocurrent in a four-terminal Hall-bar geometry may yield a photoconductor-like detection mechanism that exhibits zero net dark current, even in the presence of a large voltage bias. This unusual characteristic, when combined with the giant Hall photoconductivity of narrow-gap GDMs, provides a new route toward high-sensitivity long-wavelength detectors. 

\section*{Acknowledgments}
We are grateful to Valla Fatemi and Alex Frenzel for helpful discussions, as well as a critical reading of our manuscript. This work was supported by the Singapore National Research Foundation (NRF) under NRF fellowship award NRF-NRFF2016-05 (J.C.W.S.), and by startup funds from UW-Madison (M.A.K.)

ACS: This document is the author's version of a Submitted Work that was subsequently accepted for publication in NanoLetters, copyright American Chemical Society after peer review. To access the final edited and published work see http://dx.doi.org/10.1021/acs.nanolett.6b02559.

\setcounter{equation}{0}
\renewcommand{\theequation}{S-\arabic{equation}}
\makeatletter
\renewcommand\@biblabel[1]{S#1.}

\section{Supplementary Information to ``Giant Hall photoconductivity in narrow-gapped Dirac materials''}

In this supplement we discuss Berry curvature and Hall photoconductivity in dual-gated Bilayer graphene which can host an energy gap tunable by an inter-layer electric potential. 

\subsection{$\sigma_{xy}^{\rm pe}$ in dual-gated bilayer graphene}

For small inter-layer biases, the low energy hamiltonian of dual-gated bilayer graphene can be approximated in the compact fashion similar to Eq.~(3) of the main text as $H_{K,K'} = \vec d(\vec p) \cdot \boldsymbol{\tau}$ but with $\vec d(\vec p)$ written as~S\cite{martin,koshino,zhang}
\be
\vec d(\vec p) = (v^2 [p_x^2 - p_y^2]/\gamma_1, \, 2\zeta v^2 p_xp_y/\gamma_1, \, \Delta),  
\label{eq:blg}
\ee
where $\gamma_1$ is the nearest-interlayer-neighbor hopping amplitude, and $\zeta =\pm 1$ for electron in the $K$ and $K'$ valleys as in the main text. This hamiltonian is valid so long as $\gamma_1 \gg \Delta$, and $v|\vec p| \gg \Delta$. Since $\gamma_1 \approx 0.38 \, {\rm eV}$~S\cite{Kuzmenko}, and $2\Delta \approx {\rm few} - 100\, {\rm meV}$ have been reported~S\cite{zhang2,tarucha,zhang3,long}, Eq.~(\ref{eq:blg}) captures the low-energy physics of dual-gated bilayer graphene in the density range $n\gg \tilde{n}$ we are interested in. Here $\boldsymbol{\tau} = \tau_x \hat{\vec x} +\tau_y \hat{\vec y} + \tau_z\hat{\vec z}$, where $\tau_{x,y,z}$ are the Pauli matrices which act on the A/B sub-lattice space in alternate layers [i.e.  the spinor is written as $(\psi_{B,{\rm top}}, \psi_{A,{\rm bottom}})$]. Diagonalizing Eq.~(\ref{eq:blg}) yields energies
\be
\epsilon_\pm = \pm \sqrt{ v^4 |\vec p|^4/\gamma_1^2 + \Delta^2}.
\ee

The Berry curvature for dual-gated bilayer graphene in Eq.~(\ref{eq:blg}) can be written as~S\cite{koshino,zhang,tarucha}
\be
\Omega_{\pm}^\zeta = \pm \frac{2\zeta \hbar^2 \Delta \gamma_1 v^4 |\vec p|^2}{(v^4 |\vec p|^4 + \gamma_1^2 \Delta^2)^{3/2}}
\ee
Interestingly, while peaked near the band edges, $\Omega(\vec p)$ for dual-gated BLG vanishes for $\vec p = 0$, in contrast to Eq.~(1) in the main text for gapped single layer graphene. 

Following the same procedure as the main text, we can write the initial carrier density per spin {\it before} irradiation as $n_{K}=n_{K'} = n_0$. Using Eq.~(5) and Eq.~(6) of the main text, we obtain a photo-induced Hall conductivity in dual-gated BLG as 
\begin{align}
\sigma_{xy}^{\rm pe} & = \frac{Ne^2}{h} \Big[ \mathcal{G}_K (n_{\rm pe}^K/2) +  \mathcal{G}_{K'} (n_{\rm pe}^{K'}/2) \Big], \nonumber \\
\mathcal{G}_\zeta (x) & = \zeta \Bigg( 1 - \Big[\frac{(\tilde{n} n_1)^{1/2}}{\sqrt{x^2 +\tilde{n} n_1 }} + \frac{(\tilde{n} n_1)^{1/2}}{\sqrt{(x + n_0)^2 +\tilde{n} n_1 }}\Big] \Bigg),
\label{eq:blgsigmaxy}
\end{align}
 where $\tilde{n} = \Delta^2/(4\pi \hbar^2 v^2)$ is the same as in the main text and $n_1 = \gamma_1^2/(4\pi \hbar^2 v^2)$, and $N=2$ is the spin degeneracy. We note that the large Berry flux per valley in dual-gated bilayer graphene yields a large maximal value of $\sigma_{xy}^{\rm pe} = 2N e^2/h$. Eq.~(\ref{eq:blgsigmaxy}) yields a $\sigma_{xy}^{\rm pe}$ similar to that discussed in the main text, exhibiting a saturation behavior, and can also be tuned by carrier doping $n_0$. 

In contrast to gapped single layer graphene, the characteristic density scale in dual-gated bilayer graphene is 
\be
n_* = \sqrt{n_1 \tilde{n}}. 
\ee
We note that $n_*> \tilde{n}$ since $\gamma_1 \gg \Delta$. However, large values of $\sigma_{xy}^{\rm pe}$ can still be achieved for small gap sizes in Eq.~(\ref{eq:blgsigmaxy}), and for relatively low photoexcited carrier density. Taking $\Delta = 10\, {\rm meV}$ and $\gamma_1 = 0.38 \, {\rm eV}$, we find relatively low $n_* = 6.8 \times 10^{10} \, {\rm cm}^{-2}$. Here we have used $v = 10^8 {\rm cm} \, s^{-1}$ as in the main text.

\end{document}